\documentclass{aa}

\usepackage{amssymb}
\usepackage{amsmath}
\usepackage{graphicx}
\usepackage{verbatim}
\usepackage{color}

\begin{document}

\title{Initiation of Alfv\'enic turbulence by Alfven wave collisions: a numerical study}

\author{S.~V.~Shestov\inst{1} \and Y.~M.~Voitenko\inst{2} \and  A.~N.~Zhukov\inst{1}\fnmsep\inst{3}}

\institute{Solar-Terrestrial Centre of Excellence --- SIDC, Royal Observatory of Belgium, 
    Avenue Circulaire 3, B-1180, Brussels, Belgium;  \email{s.shestov@oma.be}
       \and
    Solar-Terrestrial Centre of Excellence, Space Physics Division, Royal Belgian Institute for Space Aeronomy, Brussels, Belgium
        \and
    Skobeltsyn Institute of Nuclear Physics, Moscow State University, Leninskie gory, 119991, Moscow, Russia  }

\date{28 September 2021}

\abstract{In the framework of compressional magnetohydrodynamics (MHD), we study numerically the commonly
accepted presumption that the Alfv\'{e}nic turbulence is generated by the
collisions between counter-propagating Alfv\'{e}n waves (AWs). In the
conditions typical for the low-beta solar corona and inner
solar wind, we launch in the three-dimensional simulation box two
counter-propagating AWs and analyze polarization and spectral properties of
perturbations generated before and after AW collisions. The observed
post-collisional perturbations have different polarization and smaller
cross-field scales than the original waves, which supports theoretical
scenarios with direct turbulent cascades. However, contrary to theoretical
expectations, the spectral transport is strongly suppressed at the scales
satisfying the classic critical balance of incompressional MHD. Instead, a
modified critical balance can be established by colliding AWs with
significantly shorter perpendicular scales. We discuss consequences of these
effects for the turbulence dynamics and turbulent heating of compressional
plasmas. In particular, solar coronal loops can be heated by the
strong turbulent cascade if the characteristic widths of the loop
sub-structures are more than 10 times smaller than the loop width. The
revealed new properties of AW collisions have to be incorporated in the
theoretical models of AW turbulence and related applications.}


\maketitle

\section{Introduction}

Recent studies have revealed that the turbulence in magnetized plasmas is
greatly affected by the Alfv\'{e}n wave effects. The well-documented example
is the solar-wind turbulence whose nature is essentially Alfv\'{e}nic and
turbulent fluctuations can be approximately described as Alfv\'{e}n waves
(AWs) \citep{doi:10.1029/JA076i016p03534,2013LRSP...10....2B}. The standard
magnetohydrodynamic (MHD) description of Alfv\'{e}nic turbulence in
astrophysical and laboratory plasmas is based on the interaction of
oppositely propagating incompressible wave packets %
\citep{1963AZh....40..742I,1965PhFl....8.1385K}.

Following significant previous work on the weak turbulence in incompressible
MHD \citep{Sridhar1994,1995ApJ...447..706M,1996ApJ...465..845N,2000JPlPh..63..447G},
the more recent work \citep{Howes2013} has described the mechanism of
turbulent energy transfer via AW collisions in more detail. The authors
showed analytically that two colliding counter-propagating AWs with
wavevectors $\mathbf{k}_{0}^{-}=k_{\perp}^{-}\mathbf{\hat{y}}+k_{\parallel }\mathbf{\hat{z}}$
and $\mathbf{k}_{0}^{+}=k_{\perp}^{+}\mathbf{\hat{x}}-k_{\parallel }\mathbf{\hat{z}}$
first produce a specific intermediate wave
with $\mathbf{k}_{2}=k_{\perp}^{+}\mathbf{\hat{x}}+k_{\perp}^{-}\mathbf{\hat{y}}$, and then its interaction with the initial waves produces the
tertiary waves with wavevectors $\mathbf{k}_{3}^{-}=k_{\perp}^{+}\mathbf{\hat{x}}
+2k_{\perp}^{-}\mathbf{\hat{y}}+k_{\parallel }\mathbf{\hat{z}}$ and 
$\mathbf{k}_{3}^{+}=2k_{\perp}^{+}\mathbf{\hat{x}}+k_{\perp}^{+}\mathbf{\hat{y}}
-k_{\parallel }\mathbf{\hat{z}}$. Here $\mathbf{\hat{x}}$, $\mathbf{\hat{y}}$ and
$\mathbf{\hat{z}}$ are the unit Cartesian vectors such that $\mathbf{\hat{z}}$
is parallel to the background magnetic field $\mathbf{B}_{0}$. These analytical results
have been confirmed by both gyrokinetic simulations in the MHD limit \citep{2013PhPl...20g2303N}
and experimentally in the laboratory 
\citep{2013PhPl...20g2901D,2014ITPS...42.2534D,2016PhPl...23b2305D}. Since
the energy is transferred to AWs with higher perpendicular wavenumbers, this
process represents an elementary step of the direct turbulent cascade in
which energy is transferred from larger to smaller scales.

\citet{Goldreich1995} introduced the critical balance conjecture and
developed their famous model of strong anisotropic MHD turbulence. The
critical balance assumes that the linear (wave-crossing) and nonlinear (eddy
turnover) times are equal at each scale. Whereas the critical balance
remains a physically reliable hypothesis not strictly derived from basic
principles, it allows for a phenomenological prediction of turbulence
properties, in particular the energy spectrum $\sim k_{\perp}^{-5/3}$ and
anisotropy of turbulent fluctuations. The Goldreich \& Sridhar model gave
rise to many important insights in the turbulence nature and resulted in
many theoretical, numerical, and experimental studies 
\citep[see e.g.][and references therein]{2018JPlPh..84a9009V, 2018JPlPh..84a9003V, Mallet2015}.
It is worth noting that the critical balance conjecture is essentially a
statement implying persistence of linear wave physics in the strongly
turbulent plasma.

Despite extended investigations of the critically balanced turbulence, many
actual problems remain open, such as the non-zero cross-helicity effects in
the presence of shear plasma flows \citep{2016Ap&SS.361..364G}, or non-local
effects in AW collisions \citep{2008ApJ...682.1070B}. Also, the plasma
compressibility can introduce surprising effects in the behavior of MHD
waves \citep{Magyar2019}.

Numerical simulations of turbulence are usually done either via
numerical codes for reduced MHD or using analytical frameworks 
\citep{Beresnyak_2014,Beresnyak_2015,Mallet2015,PhysRevResearch.2.023189},
pseudo-spectral \citep{Chandran2019} and gyrokinetic %
\citep{2018JPlPh..84a9003V}. \citet{2017PhRvE..96b3201P,2017JPlPh..83a7008P}
performed simulations using compressible MHD, Hall MHD, and hybrid
Vlasov-Maxwell codes; the 2.5D geometry used in these works did not allow to
take into account nonlinear terms $\sim \left( \mathbf{v}^{\pm }\cdot 
\mathbf{\nabla }\right) \mathbf{v}^{\mp }$ and $\sim \left( 
\mathbf{b}^{\pm }\cdot \mathbf{\nabla }\right) \mathbf{b}^{\mp }$
($v^{\pm}$ and $b^{\pm}$ are velocity and magnetic
fluctuations in $\pm$ waves) for AWs with $\mathbf{k}_{\perp}^{+}\times \mathbf{k}_{\perp}^{-}\neq 0$.

Using compressible MHD model in 3D, we study numerically the
commonly accepted presumption that the AW turbulence is generated by the
collisions between counter-propagating AWs, particularly the wavenumber
dependence of the amplitudes of induced waves. Our simulations reveal that
the AW collisions can occur in two regimes, the first one corresponding to
the case of strong turbulence which follows theoretical explanation, and the
second one corresponding to larger scales which obviously is governed by a
different mechanism.

\section{Physical and Numerical setup}

\label{setup-sec} The simulations were performed in 3D using the numerical
code MPI-AMRVAC \citep{2014ApJS..214....4P}. The code applies the Eulerian
approach for solving the compressible resistive MHD equations: 
\begin{equation}
\frac{\partial \rho }{\partial t}+\nabla \cdot (\rho \mathbf{v})=0,
\label{mass_v}
\end{equation}
\begin{equation}
\frac{\partial (\rho \mathbf{v})}{\partial t}+\nabla \cdot \left( \mathbf{v}%
\rho \mathbf{v}-\mathbf{BB}\right) +\nabla p_{\mathrm{tot}}=0,  \label{mom_v}
\end{equation}
\begin{equation}
\frac{\partial \mathbf{B}}{\partial t}+\nabla \cdot \left( \mathbf{vB}-%
\mathbf{Bv}\right) =-\nabla \times (\eta \mathbf{J}),  \label{ind_v}
\end{equation}
\begin{equation}
\frac{\partial e}{\partial t}+\nabla \cdot \left( \mathbf{v}e-\mathbf{BB}%
\cdot \mathbf{v}+\mathbf{v}p_{\mathrm{tot}}\right) =\nabla \cdot \left( 
\mathbf{B}\times \eta \mathbf{J}\right) ,
\end{equation}
where $e$, $\rho $, $\mathbf{v}$, $\mathbf{B}$ are the total energy density,
mass density, velocity, and magnetic field, $p=(\gamma -1)(e-\rho \mathbf{v}%
^{2}/2-{B}^{2}/2)$ is the thermal pressure, $p_{\mathrm{tot}}=p+{B}^{2}/2$
is the total pressure, $\mathbf{J}=\nabla \times \mathbf{B}$ is the electric
current density, $\eta $ is the electrical resistivity, and $\gamma $ is the
ratio of specific heats. The magnetic field is measured in units for which
the magnetic permeability is 1. Since in this study we are not interested in
dissipative processes, we take $\gamma =5/3$, and $\eta =0$. We used three
following normalization constants: the length $L_{\mathrm{N}}=1$~Mm, the
magnetic field $B_{\mathrm{N}}=20$~G, and the density $\rho _{\mathrm{N}%
}=1.67\times 10^{-15}$~g~cm$^{-3}$. This determined normalization for other
physical quantities: electron concentration $n_{\mathrm{N}}=10^{9}$~cm$^{-3}$%
, speed $v_{\mathrm{N}}={B_{\mathrm{N}}}/\sqrt{4\pi \rho _{N}}=1\,380$~km s$%
^{-1}$, and time $t_{\mathrm{N}}=L_{\mathrm{N}}/v_{\mathrm{N}}=0.7246$~s.

The simulations are performed in 3D in Cartesian geometry with a rectangular
numerical box. The background magnetic field $B_{0}=20$~G is directed along $%
z$-axis. Equilibrium plasma parameters are taken typical for the
solar coronal base: $n_{e}=10^{9}$~cm$^{-3}$ ($\rho _{0}=1.67\times 10^{-15}$%
~g~cm$^{-3}$) and temperature $T=1$~MK, which determines the plasma beta
parameter $\beta =0.017$. The Alfv\'{e}n speed in equilibrium plasma is $%
v_{A}=B_{0}/\sqrt{4\pi \rho _{0}}=1\,380$~km~s$^{-1}$ or $v_{A}=1$ in
normalized units, and the sound speed is $C_{S}=\sqrt{\gamma \beta /2}%
v_{A}=0.11v_{A}$.

\begin{figure}[tbp]
\includegraphics[width=9cm]{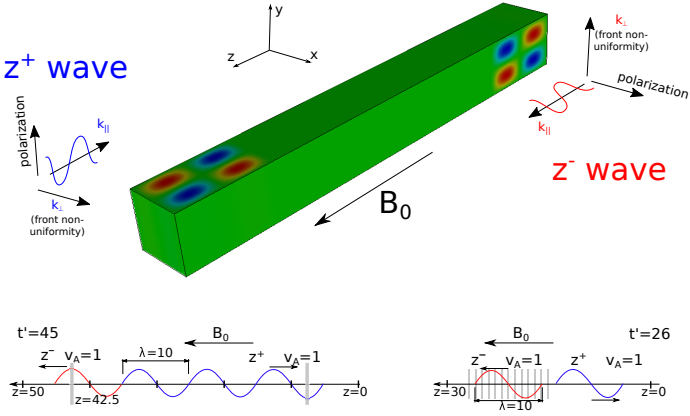}
\caption{\textit{Top}: physical setup, early phase. The green rectangle
denotes the numerical box with equilibrium plasma, the red and blue areas
represent velocity perturbations (positive and negative) of the $z^-$ wave
(far boundary) and $z^+$ wave (near boundary). Polarization planes and
non-uniformity directions are annotated. \textit{Bottom}: longitudinal sketches,
late phase after AWs collision: main setups with 1
period in $z^-$ and 3 periods in $z^+$ wave (\textit{left}), and
complimentary setups with 1 periods in both waves (\textit{right}). The
grey areas denote cross-sections taken for further analysis.}
\label{phys-setup}
\end{figure}

In order to induce counter-propagating Alfven waves, we set the components
of magnetic field and velocity at the $z$-boundaries of the simulation
volume. The forward wave propagating in $+z$ direction along $\mathbf{B}_{0}$
is initiated at $z=0$ by the following forcing: 
\begin{eqnarray}
b_{x} &=&b\sin \left( \omega t\right) \sin (k_{\perp}^{-}y); \\
v_{x} &=&-u\sin \left( \omega t\right) \sin (k_{\perp}^{-}y); \\
v_z &=&A_{p}\left[ 1-\sin \left( 2\omega t\right) \right] \sin (k_{\perp
}^{-}y),  \label{vzm}
\end{eqnarray}%
and the backward wave propagating in $-z$ direction is initiated at $z=z_{\mathrm{max}}$: 
\begin{eqnarray}
b_{y} &=&b\sin \left( \omega t\right) \sin (k_{\perp}^{+}x); \\
v_{y} &=&u\sin \left( \omega t\right) \sin (k_{\perp}^{+}x); \\
v_z &=&A_{p}\left[ -1+\sin \left( 2\omega t\right) \right] \sin (k_{\perp
}^{+}x),  \label{vz}
\end{eqnarray}%
where $\omega =k_z v_{A}=2\pi \left( \lambda_z\right) ^{-1}v_{A}$ is the
angular frequency, the parallel wavelength $\lambda_z=10$~Mm
(always the same constant in all setups), the initial amplitudes of magnetic
field $b$ and velocity $u=b/\sqrt{4\pi \rho }$ are either 3.33\% or 10\% of
$B_{0}$ and $v_{A}$, respectively, and $v_z\sim A_{p}=0.25\left(
v_{A}^{2}-C_{S}^{2}\right) ^{-1}u^{2}v_{A}$ represents the ponderomotive
component of the speed (its order is $10^{-3}$). Boundary conditions at
other boundaries are periodic. Introduction of $A_{p}\neq $ $0$ insures a
smooth solution of the MHD equations at the boundaries; its influence is
studied in Sect.~\ref{selfint-sec}. The physical configuration is shown in
Fig.~\ref{phys-setup}.

The described above forcing is applied during 1 period for the forward wave
and 3 periods for the backward wave, which we call the main setups hereafter
(see Table~\ref{tab-parameters} for setup parameters). Beside the main
setups, we run several complementary simulations without backward wave, or
with different amplitudes of counter-propagating waves, or setups with a
single period in both waves.

As suggested by the nonlinear term $\left( \mathbf{z}^{\pm }\cdot 
\mathbf{\nabla }\right) \mathbf{z}^{\mp }$ in Els\"{a}sser form of
MHD equations, in order to allow for effective interactions, the
counter-propagating AWs should have different polarizations. In our setups, 
the forward wave is polarized along $x$ and its wavevector $\mathbf{k}_{\perp}^{-}\parallel \mathbf{\hat{y}}$;
the backward wave is polarized along $y$-axis and $\mathbf{k}_{\perp}^{+}\parallel \mathbf{\hat{x}}$ (see
Fig.~\ref{phys-setup}).

\begin{table}[tbp]
\caption{Parameters of the numerical setups. }
\label{tab-parameters}{\scriptsize \ 
\begin{tabular}{l|l}
\hline\hline
\multicolumn{2}{l}{\textbf{Main setups}:} \\ \hline
\multicolumn{2}{l}{\textbf{High-resolution}} \\ 
{Numerical box} & {$256\times256\times 512$ pixels} \\ 
{$L_z$} & {50~Mm} \\ 
{$L_x$, $L_y$} & {equal to $\lambda_\perp$} \\ 
{Number of periods} & {$z^-$ -- 1 period} \\ 
{\ } & {$z^+$ -- 3 periods} \\ 
{$u$} & {0.1 (same for $z^-$ and $z^+$)} \\ 
$\lambda_z$ (or $\lambda_\parallel$) & 10~Mm \\ 
$\lambda_\perp$ & from 0.4 to 25.0 (10 configurations) \\ 
$k_\perp$ & from 15.7 to 0.25 \\ 
$k_\perp/k_\parallel$ & from 25.0 to 0.4 \\ \hline
\multicolumn{2}{l}{\textbf{Low-resolution}} \\ 
{Numerical box} & {$128\times 128\times 256$ pixels} \\ 
{$L_z$} & {50~Mm} \\ 
{$L_x$, $L_y$} & {equal to $\lambda_\perp$} \\ 
{$u$} & {0.033 (same for $z^-$ and $z^+$)} \\ 
$\lambda_z$ (or $\lambda_\parallel$) & 10~Mm \\ 
$\lambda_\perp$ & from 0.16 to 25.0 (12 configurations) \\ 
$k_\perp$ & from 39.3 to 0.25 \\ 
$k_\perp/k_\parallel$ & from 62.5 to 0.4 \\ \hline\hline
\multicolumn{2}{l}{\textbf{Non-zero cross-helicity}:} \\ \hline
{Numerical box} & {$256\times256\times 364$ pixels} \\ 
{$L_z$} & {30~Mm} \\ 
{$L_x$, $L_y$} & {equal to $\lambda_\perp$} \\ 
{Number of periods} & {$z^-$, $z^+$ -- 1 period} \\ 
{$u^-$} & {0.1} \\ 
{$u^+$} & {0.03} \\ 
$\lambda_z$ (or $\lambda_\parallel$) & 10~Mm \\ 
{$\lambda_\perp$, $k_\perp$, $k_\perp/k_\parallel$} & same as in
high-resolution main setups \\ \hline\hline
\multicolumn{2}{l}{\textbf{Perpendicular and longitudinal structure}:} \\ 
\hline
{Numerical grid} & {$256\times256\times 364$ pixels} \\ 
{$L_z$} & {30~Mm} \\ 
{$L_x$, $L_y$} & {equal to $\lambda_\perp$} \\ 
{Number of periods} & {$z^-$, $z^+$ -- 1 period} \\ 
{$u$} & {0.1 (same for $z^-$ and $z^+$)} \\ 
$\lambda_z$ (or $\lambda_\parallel$) & 10~Mm \\ 
{$\lambda_\perp$, $k_\perp$, $k_\perp/k_\parallel$} & same as in
high-resolution main setups \\ 
& 
\end{tabular}
}
\end{table}

The numerical box has physical $z$-length $L_z$ either 50~Mm (main setups)
or 30~Mm (complementary setups). The sizes along $x$ and $y$ are set equal
to the perpendicular wavelength $\lambda_{\perp}$ (hence change from setup
to setup). The numerical box for the main setups has either $256\times
256\times 512$ pixels (high-resolution) or $128\times 128\times 256$ pixels
(low-resolution). We have verified that the decrease of numerical resolution
does affect the results: the waves start to decay during their propagation
and the wave profiles get distorted. However, this effect is small even for
the case of low-resolution setups. In complimentary setups the numerical box
always has $256\times 256\times 384$ pixels, thus its spatial resolution
coincides with that of the high-resolution setups. We compared various
numerical schemes and parameters of MPI-AMRVAC and chose the best settings (%
%
%
\verb|powel| scheme for the $\mathbf{\nabla \cdot B}$ corrector,
high-resolution numerical box etc.). We also pay special attention to
distinguish the physical phenomena from numerical artifacts.

\section{Results}

\subsection{Nonlinear effects in a single AW}

\label{selfint-sec} 
\begin{figure}[tbp]
\includegraphics[width=9cm]{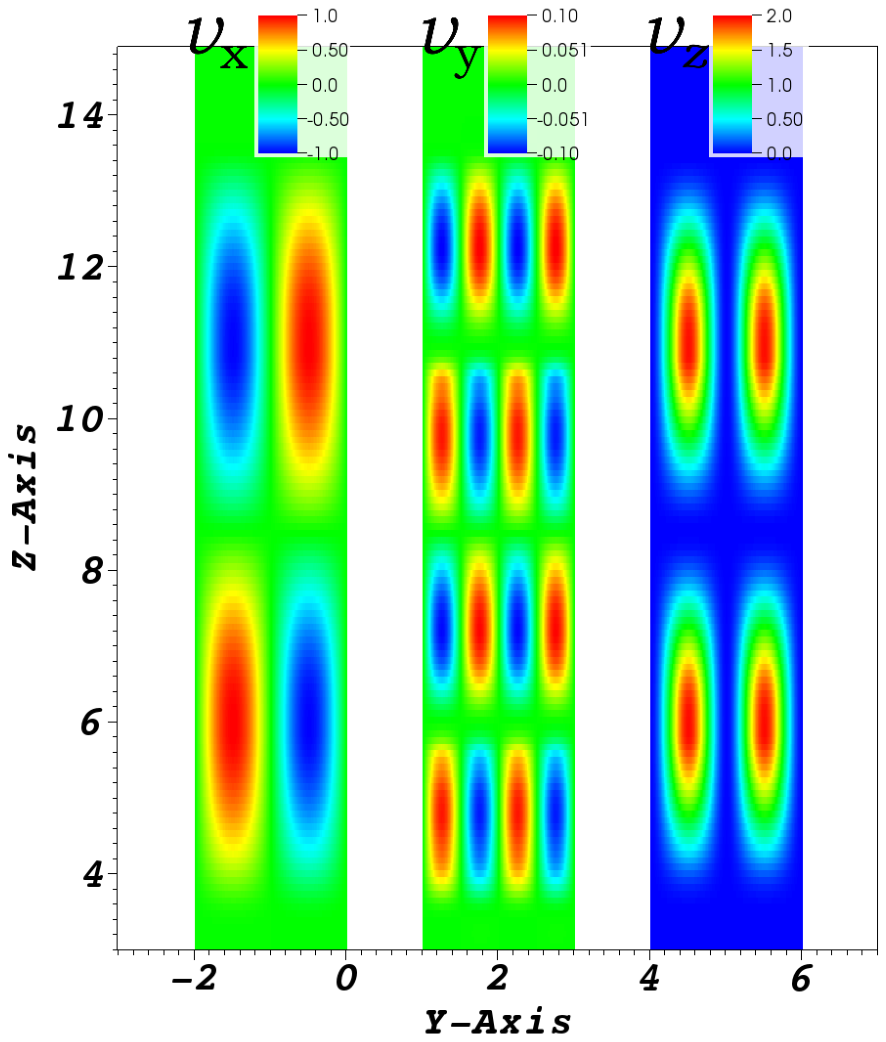}
\caption{Velocities $v_{x}$ (\textit{left}), $v_{y}$ (\textit{middle}), and 
$v_z$ (\textit{right}) in a single Alfv\'{e}n wave. The color
tables (inlines in top) have different amplitudes in different panels to
reflect the range of velocities}. Initially in the Alfv\'{e}n wave only $v_{x}$
and $B_{x}$ (not shown) and $v_z$ are driven; the $v_y$ component of
the velocity is generated self-consistently due to nonlinear
self-interaction within the Alfv\'{e}n wave.
\label{selfint}
\end{figure}

First we verify the effect of nonlinear self-interaction
within a single Alfv\'{e}n wave. In Fig.~\ref{selfint} we show the
longitudinal (along $z$) profiles of $v_{x}$, $v_{y}$, and $v_z$ of the
forward Alfv\'{e}n wave, initiated via boundary conditions described
in Sect.~\ref{setup-sec}, during its developed phase, but before the
collision with the backward wave. For visualization the quantities are
normalized by the following constants: the mother wave $v_{x}$ by the
initial amplitude $u=0.10$, the horizontal component $v_{y}$ and the
ponderomotive component $v_z$ by $A_{p}=2.53\cdot 10^{-3}$. 

The amplitude and spatial structure of the ponderomotive component $v_z$
perfectly reproduces theoretical predictions: its wavenumbers are two times
larger than in the mother wave and its amplitude varies from 0 to 2 %
\citep{McLaughlin2011,Zheng2016}. We also observed a self-consistent
generation of $v_{y}$ that appears only in oblique waves with $\lambda
_{\perp}\neq 0$ ($v_{y}=0$ at $\lambda_{\perp}=0$). Our preliminary
simulations (two-dimensional setups were sufficient there) have shown the
following trend in the variation of $v_{y}$ with varying cross-field
wavelength: the amplitude of $v_{y}$ grows proportionally to $1/\lambda_{\perp}$
at the larger scales $\lambda_{\perp}>\lambda_z$, this growth slows down at 
$\lambda_{\perp}\sim \lambda_z$, and eventually $v_{y}$ becomes a constant
independ on $\lambda_{\perp}$ at smaller scales $\lambda_{\perp}\ll \lambda_z$. 
The spatial extension of $v_{y}$ in both parallel and perpendicular directions is two times shorter than of the
mother wave. The amplitude of $v_{y}$ is always smaller than that of $v_z$.
Similar perturbations of perpendicular velocity were observed also in
torsional waves \citep{2017ApJ...840...64S}.

The observed perturbations of $v_{y}$ and $v_z$ propagate along the
magnetic field with the Alfv\'{e}n speed $v_{A}$ and are natural companions
of AWs not caused by the numerical effects or boundary conditions for $v_z$.
The perturbations always develop in AWs regardless of the ways how the
waves are initiated -- by boundary or initial conditions, with or without
boundary perturbations given by Eqs.~\ref{vzm} and \ref{vz}. In
other words, the observed propagating wave is the eigenmode of the
compressible nonlinear MHD.

We thus observe typical characteristics of AWs before they collide.

\subsection{AWs collision}

\begin{figure}[tbp]
\includegraphics[width=9cm]{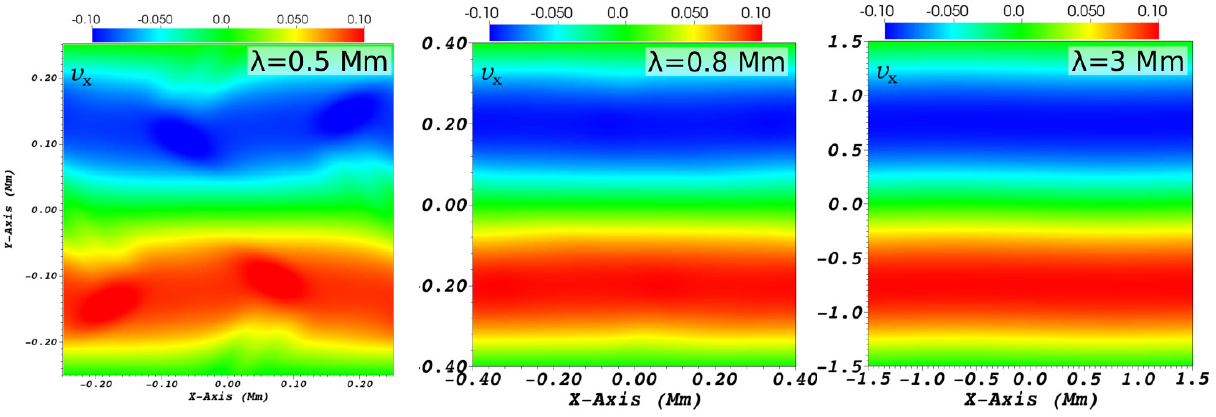}
\caption{Comparison of perpendicular profiles of $v_x$ in the leading
maximum of the $z^-$ wave in setups with $\protect\lambda_\perp = 0.5$ (%
\textit{left}), $0.8$ (\textit{middle}), and $3.0$~Mm (\textit{right}). The
axes are in Mm and are different in every panel.}
\label{2D-profiles}
\end{figure}

To study effects of the AW collisions, we let the two
counter-propagating waves to fully propagate through each other, and
analyze perpendicular profiles of $v_{x}$ of the forward-propagating $z^{-}$
wave in its leading maximum -- $x-y$ plane with $z=42.5$ at instant $t=45$,
see Fig.~\ref{phys-setup}, bottom left panel (main setups with $u=0.1$ are
used). In Fig.~\ref{2D-profiles} the panels show $v_{x}$ of three different
setups with $\lambda_{\perp}=0.5$, $0.8$, and $3.0$~Mm. The perturbations
of the wave profiles depend on the perpendicular scale: they are significant
for smallest $\lambda_{\perp}=0.5$, moderate for $\lambda_{\perp}=0.8$,
and weak for the largest $\lambda_{\perp}=3.0$. Similar perturbations are
also observed in the $z^{+}$ wave. However, in setups with only one wave
present, such perturbations do not appear, and hence their development can
be attributed to AWs collision.

Appearance of such small-scale perturbations propagating with Alfv\'{e}n
velocity can be treated as generation of new AWs at smaller perpendicular
scales $\lambda_{\perp}^{\prime }<\lambda_{\perp}$.

\subsection{Dependence on perpendicular scales}

\begin{figure}[tbp]
\includegraphics[width=9cm]{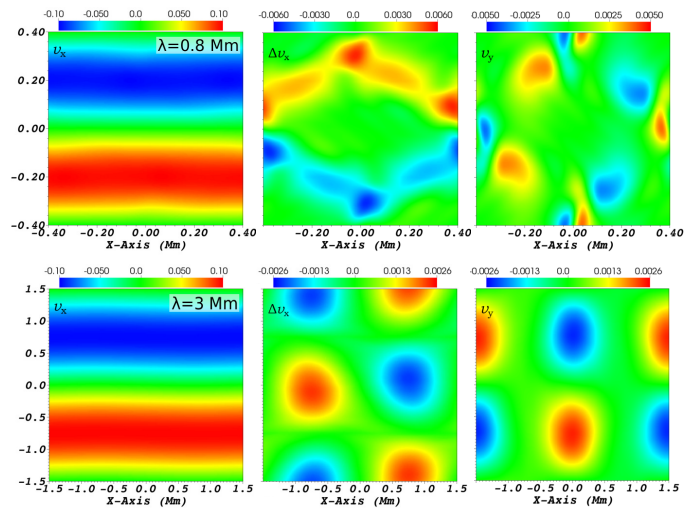}
\caption{Perpendicular profiles of velocities in setups with $\lambda_\perp=0.8$
(\textit{top}) and $\lambda_\perp=3$ (\textit{bottom}). 
\textit{Left}: measured $v_x$; \textit{middle}: difference $\Delta v_x$
between the measured $v_x$ and a harmonic function; \textit{right}: measured 
$v_y$. Perturbations of $\Delta v_x$ and $v_y$ are produced in result of AWs
collision. In each panel the color table matches the maximum amplitude of
the measured quantity.}
\label{Data2_perpendicular}
\end{figure}

In order to distinguish the nonlinearly generated waves from the mother
wave, we further analyze the wave profiles in the perpendicular
cross-section of $z^{-}$ wave. We extract the perturbed velocity $\Delta
v_{x}$ by subtracting the initial harmonic profile of $v_{x}$, $\Delta
v_{x}=v_{x}-A\sin (k_{\perp}^{-}y)$, where the amplitude $A$ is adjusted to
cancel the perturbation in the wave maximum. The results are shown in Fig.~%
\ref{Data2_perpendicular} for the setups with $\lambda_{\perp} =0.8$ (top
panels) and $\lambda_{\perp}=3.0$ (bottom panels). The left panels show $%
v_{x}$, middle $\Delta v_{x}$, and right $v_{y}$. The induced velocities $%
\Delta v_{x}$ and $v_{y}$ have amplitudes $\sim (0.02\div 0.03)u$ and are
non-uniform in both $x$ and $y$ directions.

To evaluate numerical effects, we made the similar analysis for $z^{-}$ wave
in the absence of $z^{+}$ waves. Here the perturbations $\Delta v_{x}$ are
observed as well; but they have at least factor 10 smaller amplitude and are
uniform along $x$. It means that numerical effects produce significantly
weaker perturbations with different spatial profiles. On the contrary, after
collisions with counter-propagating $z^{+}$ waves, the perturbations
co-propagating with $z^{-}$ waves have both $v_{x}$ and $v_{y}$ components,
larger amplitudes, and profiles non-uniform both along $y$ and $x$, which
cannot be ascribed to numerical effects. Furthermore, the perturbations of $%
v_{y}$ generated by the AW collisions can not be attributed solely to the
single AW self-interaction where perturbations of the $v_{y}$ are zero at
the original wave maximum.

The spatial patterns of the induced velocities fall in two distinct groups:
all spatial patterns at $\lambda_{\perp}<\lambda_{\perp}^{tr}$ are
similar to that shown on the top panels in Fig.~\ref{Data2_perpendicular},
and all patterns at $\lambda_{\perp}\geq \lambda_{\perp}^{tr}$ are
similar to that shown on the bottom panels (the transition scale $\lambda
_{\perp}^{tr}=3.0$ for $u=0.1$ used in this figure). The perturbations in
the former group have a current-sheet structuring, similar to that reported
by \citet{2018JPlPh..84a9003V} for the strong turbulence regime. The
perturbations in the second group have symmetric structure. The same two
groups of spatial structures are also observed in the setups with different
amplitudes $u$, but with different transition scales, such that $\lambda_{\perp}^{tr}$
is larger for smaller $u$ (for example, $\lambda_{\perp}^{tr}=4.0$ for $u=0.033$).

\begin{figure}[tbp]
\includegraphics{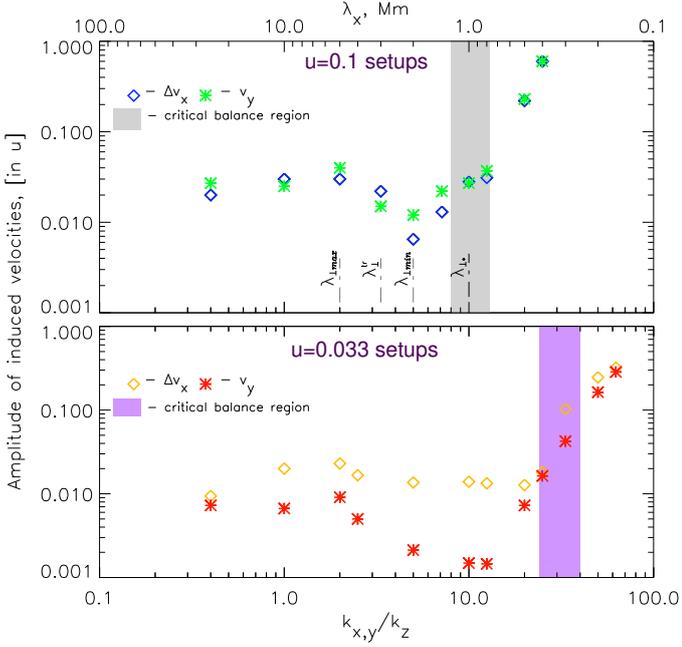}
\caption{Dependence of the amplitudes of the induced waves on $k_{\perp
}/k_{\parallel }$ of the original wave for the $u=0.1$ (\textit{top}) and
$u=0.033$ (\textit{bottom}) setups. Diamonds and asterisks correspond to the
$\Delta v_{x}$ and $v_{y}$ respectively.
Filled areas correspond to the regions of classic critical balance in
incompressible MHD calculated for particular $u$. In the top panel
the values of $\protect\lambda_{\perp\mathrm{max}}$, $\protect\lambda_{\perp%
\mathrm{min}}$, $\protect\lambda_{\perp}^{tr}$, and $\protect\lambda_{\perp *}$ are shown.}
\label{fig-eff}
\end{figure}

The dependence of the amplitudes of induced waves on the perpendicular
scales is shown in Fig.~\ref{fig-eff}. The diamonds correspond to $\Delta
v_{x}$ and asterisks correspond to $v_{y}$. For $u=0.1$ the symbols are blue
and green, for $u=0.033$ they are orange and red. Gray and pink regions
indicate the wavenumber ranges where the wave collisions should generate the
strong (critically balanced) turbulence with $k_{\perp}/k_{\parallel }\sim
v_{A}/u$ for $u=0.1$ and $u=0.033$, respectively. In both these cases the
amplitude behavior is similar. At largest $\lambda_{\perp}$ the amplitudes
of the induced waves are much smaller than the amplitudes of the original
waves and the resulting AW turbulence should be weak. As $\lambda_{\perp}$
decreases, the induced amplitudes first increase slowly and reach a maximum.
This maximum is still much smaller than the initial AW amplitude and is
reached at $\lambda_{\perp}=\lambda_{\perp \mathrm{max}}$ that is still
much larger than the perpendicular scale given by the critically balance
condition, $\lambda_{\perp \mathrm{max}}\gg $ $\lambda_{\perp \ast
}=\lambda_{\parallel }u/v_{A}$ ($\lambda_{\perp \mathrm{max}}$, 
$\lambda_{\perp \ast}$ and other characteristic perpendicular scales are
shown in Fig.~\ref{fig-eff}). When $\lambda_{\perp}$ decreases further beyond
$\lambda_{\perp \mathrm{max}}$, the induced amplitudes decrease and reach a
minimum at $\lambda_{\perp}=$ $\lambda_{\perp \mathrm{min}}$ that is
still larger than $\lambda_{\perp \ast }$. After this minimum, a strong
increase of induced perturbations occurs in the region where 
$\lambda_{\perp}$ becomes several times shorter than $\lambda_{\perp \ast}$.
Amplitudes of generated perturbations become there comparable to the
amplitudes of initial waves and such collisions can generate strong
turbulence.

While the observed strengthening of the nonlinear interaction with
decreasing $\lambda_{\perp}$ is expected taking into account that the
responsible nonlinear term is $\sim \left( \mathbf{z}\cdot \nabla \right) 
\mathbf{z\sim } $ $\lambda_{\perp}^{-1}$, the depression observed at
$\lambda_{\perp}\gtrsim $ $\lambda_{\perp \ast }$ and its influence on the
transition from weak to strong turbulence need further investigations. At
present we can only state that this depression should result in a shift of
the weak-strong turbulence transition to the perpendicular scales
significantly shorter than that prescribed by the standard critical balance
condition.

\subsection{Influence of several collisions}

\begin{figure}[tph]
\includegraphics[width=9cm]{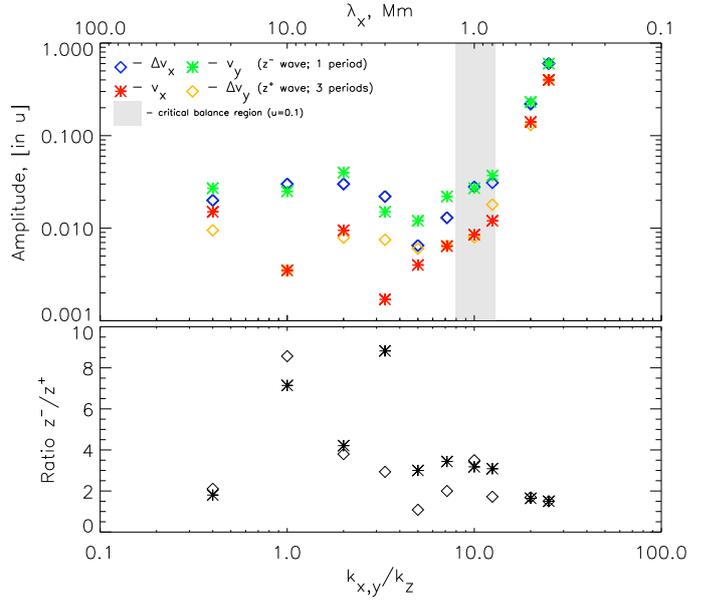}
\caption{Dependence of the amplitudes of induced waves on the wavenumber
ratio for different number of wave collisions ($z^{-}$ wave collides 3 times
and $z^{+}$ wave collides once). \textit{Top}: amplitudes of the
perturbations. \textit{Bottom}: ratio of the amplitudes $z^{-}/z^{+}$,
diamonds correspond to original polarization ($v_x$ in $z^-$, and $v_y$ in 
$z^+$) and asterisks correspond to perpendicular polarization.}
\label{fig-zp-zm}
\end{figure}

Since the initiated $z^{-}$ and $z^{+}$ waves contain 1 and 3 periods,
respectively, the $z^{-}$ wave can interact with 3 periods of the
counter-propagating wave, whereas each period of the $z^{+}$ wave can
interact with only one period of $z^{-}$. We thus expect different
amplitudes of the induced perturbation propagating in $z^{-}$ and $z^{+}$
directions. To verify this, we measure the perturbations accompanying the $%
z^{+}$ wave using the same technique as for $z^{-}$ wave (remember that in $%
z^{+}$ wave the roles of $v_{x}$ and $v_{y}$ are exchanged). Comparison of
the corresponding perturbations in the $z^{-}$ and $z^{+}$ waves is given in
Fig.~\ref{fig-zp-zm}. The top panel shows the amplitudes of perturbations
accompanying $z^{-}$ (blue and green symbols) and $z^{+}$ (orange and red
symbols), the bottom panel shows the ratio $z^{-}/z^{+}$ of the
perturbations with the corresponding (orthogonal) polarizations. In both
panels the diamonds denote perturbations with the same polarization as in
the original waves ($\Delta v_{x}$ in $z^{-}$, $\Delta v_{y}$ in $z^{+}$),
and the asterisks denote the complimentary polarization.

The behavior of $z^{+}$ perturbations as function of $\lambda_{\perp}$ is
qualitatively similar to that of $z^{-}$ perturbations. At smallest $\lambda
_{\perp}$ the ratio of the $-/+$ perturbations is about 2, then approaches
3 with the scale increase, then increases significantly at $\lambda_{\perp
}/\lambda_{\parallel }\sim 1$, and finally drops again to 2 at large
perpendicular scales $\lambda_{\perp}/\lambda_{\parallel }>1$. In the
region of (super-)strong turbulence the observed ratio $z^{-}/z^{+}<3$ means
inapplicability of the perturbation theory: already after the first
interaction the wave profiles are distorted significantly and the following
collisions do not add much.

\subsection{Non-zero cross-helicity case}

In this section we analyze the effects of non-zero cross-helicity
(imbalance) when the counter-propagating initial waves have different
amplitudes. This situation is common in the fast solar wind 
\citep{1990GeoRL..17..283T, 1998ApJ...507..984L} and also occurs in
numerical simulations in local subdomains of the simulation box 
\citep{2009PhRvL.102b5003P}.

\begin{figure}[htp]
\includegraphics[width=9cm]{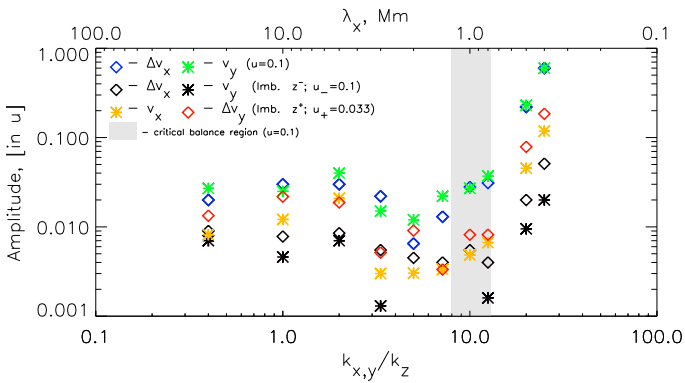}
\caption{Amplitude of the induced waves for the case of non-zero
cross-helicity: $u^-=0.1$, $u^+=0.033$, each wave has single period. Black
symbols denote perturbations in the $z^-$ wave, orange and red symbols
denote perturbations in the $z^+$ wave. Blue and green symbols represent the
main setups ($u=0.1$, 1 and 3 periods respectively). }
\label{fig-eff-ub}
\end{figure}

We run dedicated setups with initial amplitudes $u^-=0.1$ in $z^{-}$ wave
and $u^+=0.033$ in $z^{+}$ wave, both waves have one period. We compare
measured perturbations with our main setups in Fig.~\ref{fig-eff-ub}. The
black symbols denote $z^{-}$ perturbation and orange and red symbols denote
$z^{+}$ perturbations in imbalanced setups, and blue and green symbols denote
main setups ($u=0.1$, 1 period in $z^{-}$ wave and 3 periods in $z^{+}$ wave).

The perturbations observed in imbalanced cases are smaller then in the main
setups. At the same time the perturbations (expressed in initial amplitudes $u$)
in the $z^-$ wave are $\sim 3$ times smaller then in the $z^+$ wave.

\subsection{Perpendicular Fourier spectra}

\begin{figure}[htp]
\includegraphics[width=9cm]{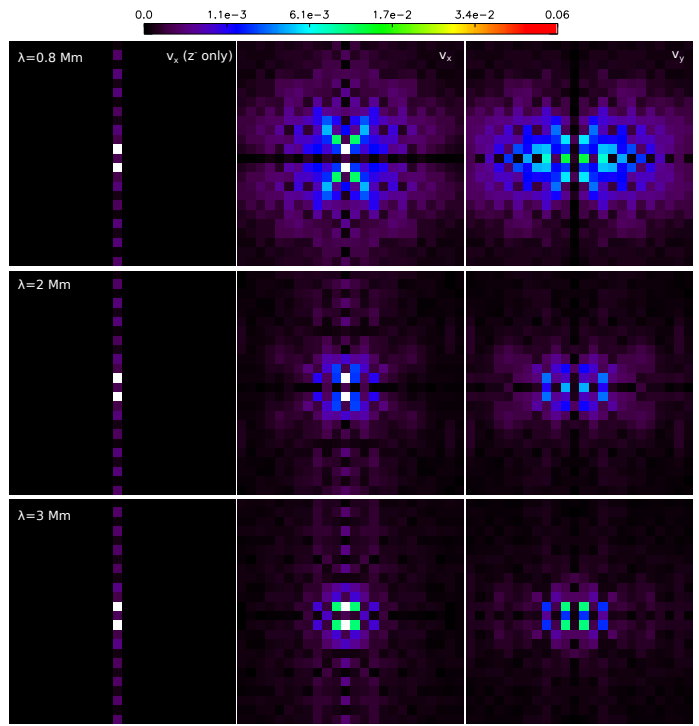}
\caption{Perpendicular Fourier spectra of $v_x$ and $v_y$ velocities
measured in the leading maximum of the $z^-$ wave. \textit{Top row}: $\lambda_\perp=0.8$;
\textit{middle row}: $\lambda_\perp=2$; 
\textit{bottom row}: $\lambda_\perp=3$. \textit{Left}: $v_x$ of the
setup with the $z^-$ wave only; \textit{middle} and \textit{right} denote $v_x$
and $v_y$ of the setups after AWs collision. In each panel the $(x,y)$-coordinates
represent corresponding Fourier wavenumbers and the color
shows intensity of a given spectral component. The quasi-logarithmic color
table is normalized to the amplitude of a harmonic wave.}
\label{spatial_spectra}
\end{figure}

In order to understand the spectral transport generated by the AW
collisions, we analyze the spatial Fourier spectra of the induced waves. The
spectra of the $v_{x}$ and $v_{y}$ velocities at the leading maximum of $z^{-}$
are given in Fig.~\ref{spatial_spectra} for $\lambda_{\perp}=0.8$ in
the top row, $\lambda_{\perp}=2.0$ in the middle row, and $\lambda_{\perp}=3.0$
in the bottom row. On the left panels, the spectra of $v_{x}$ in a
single $z^{-}$ wave are shown, on the middle and right panels the spectra of 
$v_{x}$ and $v_{y}$ after the AW collision are shown. In each panel the $(x,y)$-coordinates
represent corresponding Fourier wavenumbers and the color
shows intensity of a given spectral component. The quasi-logarithmic color
scale is normalized to the intensity of an ideal harmonic function $u\sin
(k_{\perp}^{-}y)$. This function would have only two peaks with spectral
coordinates $(0,\pm 1)$ that correspond to the brightest components in the
left and middle panels. In what follows, we will drop the $\pm$ sign
keeping in mind the inherent symmetry.

The higher-wavenumber spectral components $(0,\left\vert y\right\vert >1)$
accompanying the single $z^{-}$ wave without collisions (left panels) are
due to numerical effects. Note the low level of these components and their
uniform distribution. On the contrary, the real spectral components with
higher wavenumbers are generated by the AW collisions (middle and right
panels).

The strongest induced components at $\lambda_{\perp}=0.8$ have spectral
coordinates $(1,2)$ corresponding to the perpendicular wavevector 
$\mathbf{k_{\perp}}=k_{\perp}^{+}\mathbf{\hat{x}}+2k_{\perp}^{-}\mathbf{\hat{y}}$.
Generation of waves with such wavevectors supports the mechanism proposed by 
\citet{Howes2013} (see their Fig.~2 explaining appearance of such
``tertiary'' waves). This mechanism is summarized in the Introduction.

The same Fourier components $(1,2)$ of $v_{x}$ are also seen in the middle
row Fig.~\ref{spatial_spectra} in the case of intermediate scale $\lambda
_{\perp}=2$; in addition, the spectral components of $v_{x}$ with
coordinates $(1,1)$ corresponding to $\mathbf{k_{\perp}}=k_{\perp}^{+} 
\mathbf{\hat{x}}+k_{\perp}^{-}\mathbf{\hat{y}}$ are significant as well.

The spatial spectra of $v_{x}$ and $v_{y}$ at the largest scale $\lambda
_{\perp}=3$ are qualitatively different: the strongest induced components
have coordinates $(1,1)$ while the others are negligible. These spectral
components might be formed by a different mechanism than in the $\lambda
_{\perp}=0.8$ case.

In general, the spectral dynamics observed in our simulations, i.e.
generation of higher-wavenumber spectral components, supports scenarios with
direct turbulent cascades generated by AW collisions.

\subsection{Field-aligned structure of the induced Alfv\'en waves}

Longitudinal behavior of the Fourier components of the induced Alfv\'en
waves is studied using the following approach: we Fourier-analyze
perpendicular cross-sections at multiple $z$-coordinates, covering the
distance of slightly more then one full wavelength $\lambda_\parallel$ along 
$z$ (see bottom right sketch in Fig.~\ref{phys-setup}). In Fig.~\ref%
{longitudinal_spectra} we show longitudinal behaviour of the spectral
components of $v_x$ with coordinates $(0,1)$, $(1,1)$, $(1,2)$, $(2,1)$, and 
$(2,2)$ with different colors. The mother wave with spectral coordinates $(0,1)$
is shown with black. The top panel shows the setups with $\lambda_\perp = 0.5$,
and the bottom panel shows the setup with $\lambda_\perp = 2.0$. The intensity of
the spectral components is multiplied by factor 10 in the top panel, and by factor 100 in the bottom panel.

We observe drastically different behavior of the spectral components in
different setups. While we do not see any regularity in the larger-scale
setup, in the setup with $\lambda_{\perp}=0.5$ the growth of $(1,1)$ and $(1,2)$
components is highly correlated and their parallel scales are somehow
shorter than in initial AWs. In addition, the energy of the induced waves
tend to concentrate near the center of the mother wave.

\begin{figure}[htp]
\includegraphics[width=9cm]{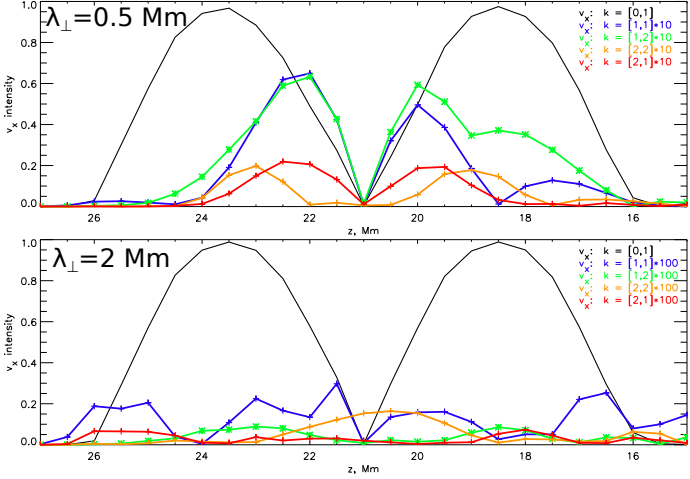}
\caption{Longitudinal dependence of amplitudes of spatial Fourier components
of $v_x$. \textit{Top}: setup with $\lambda_\perp=0.5$; \textit{bottom}: setup with
$\lambda_\perp=2$~Mm. Different colors correspond to particular spectral coordinates.}
\label{longitudinal_spectra}
\end{figure}

\section{Discussion and Application}

Results of our simulations revealed several new properties of AW collisions
in compressional plasmas, which can affect Alfv\'{e}nic turbulence and
anisotropic energy deposition in plasma species. The most striking new
property is the modified relation between the parallel and perpendicular
scales in the strong turbulence regime where the energy is efficiently
transferred to the smaller scale during one collision.

The turbulence strength is usually characterized by the nonlinearity
parameter $\chi_k \equiv \tau_k^\mathrm{L}/\tau_k^\mathrm{NL} =
\left( k_\perp v_k\right) / \left( k_z v_A\right) $, where $\tau_k^\mathrm{NL}=\lambda_{\perp}/v_k=2\pi /\left( k_{\perp}v_k\right) $
is the nonlinear mixing time, $\tau_k^\mathrm{L}=\lambda_z/v_A=2\pi /\left( k_z v_A\right)$
 is the linear (correlation) crossing
time of colliding AWs, and $v_k$ is the velocity amplitude of the
colliding AWs. Denote by $\delta v_k$ the velocity amplitude of generated
waves. When the classic critical balance condition of incompressible MHD is
satisfied, 
\begin{equation}
\chi_k=1,  \label{old}
\end{equation}%
the nonlinear mixing becomes as fast as the linear crossing and the
turbulence is believed to be strong, $\delta v_k/v_k\sim 1$ %
\citep{Goldreich1995}.

However, as follows from our simulations (see e.g. Fig.~\ref{fig-eff} showing $\delta
v_k/v_k$ as function of $k_{\perp}/k_z$ for two fixed amplitudes,
$u\equiv v_k/v_{A}=0.1$ and $0.033$, and $\delta v_k=\sqrt{\Delta
v_{x}^{2}+v_{y}^{2}}$), the spectral transport in compressible MHD is
strongly, about one order of magnitude, suppressed at $k_{\perp}/k_z$
satisfying Eq.~\ref{old}. Namely, $\delta v_k/v_k \ll 1$
at $k_{\perp}/k_z=10$ for $u=0.1$ and at $k_{\perp}/k_z=30$ for $u=0.033$.
At $k_{\perp}/k_z$ increasing further, the spectral transport eventually becomes
fast and the turbulence strong, $\delta v_k/v_k\sim 1$, which happens
at $k_{\perp}/k_z$ obeying the modified critical balance condition 
\begin{equation}
\tilde{\chi}_k=\alpha \chi _k=1,  \label{new}
\end{equation}%
where $\alpha <1$ is the factor reducing efficiency of the nonlinear mixing 
(in other words, the effective nonlinear time increases by the
factor $1/\alpha$). Consequently, the turbulence becomes strong
at perpendicular wavenumbers that are larger than in the classic critically
balanced case.

The origin and nature of $\alpha $ need further clarification. Since $\alpha
\neq 1$ arises when the plasma compressibility is taken into account, it
should depend on the relative content of thermal energy, e.g. on the plasma
$\beta$. For parameters adopted in our simulations, $\alpha \approx 0.3$ in
the critically-balanced state where the scale ratio $k_{\perp}/k_z$ obeys 
$\tilde{\chi}_k\sim 1$. Such departure from the classic critical balance
affects dynamics of the strong AW turbulence (see below). In the general
case of arbitrary scales the functional dependence $\alpha =\alpha \left(
\beta ,v_k,k_{\perp}/k_z\right) $ is complex; in particular, $\delta
v_k/v_k$ becomes a decreasing function of $k_{\perp}/k_z$ in some
interval (in Fig.~\ref{fig-eff} it happens at $k_{\perp}/k_z\lesssim 10$), which
should greatly affect the weak AW turbulence. We do not exclude that $\alpha 
$ may also depend on other plasma/wave parameters.

Let us consider the Alfv\'{e}nic turbulence driven by the fluctuating
velocity $v_{k0}$ at the wavenumber ratio $k_{\perp 0}/k_{z0}$ obeying the
critical balance condition $\tilde{\chi}_{k0}=1$, in which case the
turbulence is already strong at the driving scales. The spectral energy flux
in the inertial range is 
\begin{equation}
\epsilon_{s}=\frac{\rho v_k^{2}}{\tau_k^{\mathrm{L}}}=\alpha \frac{\rho
v_k^{2}}{\tau_k^{\mathrm{NL}}}\approx \alpha \frac{\rho
v_k^{3}k_{\perp}}{2\pi }=\mathrm{const}\equiv \alpha_{0}\frac{\rho
v_{k0}^{3}k_{\perp 0}}{2\pi },  \label{flux strong}
\end{equation}%
where $\tau_k^{\mathrm{L}}=\lambda_z/v_{A}$ is the AW collision time and 
$\rho \approx n_{0}m_{i}$ is the mass density. Note that the spectral flux
$\epsilon_s$ from Eq.~\ref{flux strong} is reduced as compared to the
incompressible strong turbulence driven at the same perpendicular scale, but
remains the same for the turbulence driven at the same parallel scale.

Assume that there is a weak dependence $\alpha =\alpha_{0}
\left(v_k/v_{k0}\right) ^{\delta }$, where $0<\delta <3/4$. Such
dependence is suggested by the following semi-empirical considerations. As
the observed spectra are power-law, the scaling of $\alpha$ with 
$v_k$ should be power law as well. Furthermore, the index $\delta $ of the
power-law dependence should be small positive to
reproduce the observed in simulations mismatch between the classic and real
critical balances (which is larger for larger wave amplitude). Moreover,
such positive values of $\delta $ appear to be compatible with the
observed spectral indexes of turbulence in the quasi-stationary solar wind,
which are slightly larger than $-5/3$ (up to $-3/2$).

The kinetic energy spectrum is then flatter than the Kolmogorov one, 
\begin{equation}
W_{s\perp }\sim v_k^{2}/k_{\perp}\sim k_{\perp}^{-5/3+2\delta /9},
\label{spectr strong}
\end{equation}%
and its spectral index varies between $-5/3$ and $-3/2$, as is typically
observed in the solar wind turbulence. In the case of $\alpha $ constant
along the critical balance path, $\delta =0$, the spectrum reduces to the
classic Kolmogorov $W_{s\perp }\sim v_k^{2}/k_{\perp}\sim k_{\perp
}^{-5/3}.$ The parallel wavenumber spectrum is, as usual, $W_{sz}\sim
v_k^{2}/k_z\sim k_z^{-2}$.

If the turbulence is weak at injection, $\tilde{\chi}_{k0}=\delta
v_{k0}/v_{k0}<1$, the cascade time increases from the strong turbulence
value $\tau_k^{\mathrm{TC}}\sim \tau _k^{\mathrm{L}}$ to the weak
turbulence value $\tau_k^{\mathrm{TC}}\sim \left( \delta
v_k/v_k\right)^{2}\tau_k^{\mathrm{L}}$. The resulting weakly
turbulent energy flux $\epsilon_{w}$ decreases as compared to the strongly
turbulent energy flux (\ref{flux strong}), $\epsilon_{w}=\tilde{\chi}_k\epsilon_{s}$: 
\begin{equation}
\epsilon _{w}=\frac{\rho v_k^{2}}{\tau _k^{\mathrm{TC}}}=\frac{\rho
v_k^{2}}{\tau _k^{\mathrm{L}}}\tilde{\chi}_k^{-2}=\mathrm{const}\equiv 
\frac{\rho v_{k0}^{2}}{\tau _{k0}^{\mathrm{L}}}\tilde{\chi}_{k0}^{-2}.
\label{flux weak}
\end{equation}%
The weakly turbulent spectrum is problematic to calculate because of a
complex dependence of $\delta v_k/v_k$ upon $k_{\perp}$ and $v_k$
(see Fig.~\ref{fig-eff}), which is unknown and difficult to guess. At present, we can
only note that the strength $\tilde{\chi}_{k0}$ of the compressional weak
turbulence is much (about one order, as is demonstrated by Fig.~\ref{fig-eff}) smaller
than the incompressional one, $\tilde{\chi}_{k0}\ll \chi_{k0}$, which
drastically decreases the weakly turbulent energy flux.

Although the large-scale MHD AWs do not dissipate directly, the turbulent
cascade transfers their energy to small scales where dissipative effects
come into play heating plasma. MHD Alfv\'{e}nic turbulence has been employed
as the mechanism for plasma heating in the solar corona and solar wind, both
from the theoretical/modeling perspective %
\citep{VanBallegooijen2011,Verdini2010} and based on experimental
observations of quiescent \citep{Morton2016,DeMoortel2014,Xie2017} and
flaring loops \citep{Doschek2014,Kontar2017}. Here we discuss how the new
properties of AW collisions observed in our simulations can affect models of
quasi-steady turbulent plasma heating in coronal loops.

Recently, \citet{Xie2017} analyzed as many as 50 loops in active regions
using observations of Extreme-ultraviolet Imaging Spectrometer (EIS) 
\citep{2007SoPh..243...19C} on board the \textit{Hinode} satellite. They
observed non-thermal widths of spectral lines and found corresponding
non-thermal velocities in the range $v_\mathrm{nt}=30\div 40$ km~s$^{-1}$,
magnetic field in the loop apexes up to 30~G, loop widths $L_{\perp}\sim
2\div 4$~Mm and loop lengths $L_z\sim 100$~Mm. \citet{2016ApJ...820...63B}
also used spectroscopic data from EIS and evaluated non-thermal velocities
in loops in 15 active regions. The typical values were somewhat smaller,
with typical values $v_{\mathrm{nt}}\sim 20$~km~s$^{-1}$; the authors
however did not provide any other parameters. Furthermore, \citet{Gupta2019}
analyzed non-thermal widths of spectral lines in high coronal loops (with
heights up to $1.4R_{\odot }$) measured by EIS and found the non-thermal
velocities in the range $20\div 30$~km~s$^{-1}$. The above values can be
used to evaluate the turbulent heating of coronal loops.

We assume that there are AW sources at the loop footpoints. These source can
be due to magnetic reconnection and/or photospheric motion (we will not
specify their origin in more details here). The perpendicular AW wavelengths 
$\lambda_{\perp 0}$ are limited by the cross-$\mathbf{B}_{0}$ scale 
$l_{\perp}$ of density filaments comprising the loops, $\lambda_{\perp
0}\lesssim l_{\perp}$ (wavenumber $k_{\perp 0}\gtrsim 2\pi /l_{\perp}$).
Note that $l_{\perp}$ can be significantly smaller than the visible loop
width $r$. On the contrary, the coronal plasma is quite homogeneous along
$\mathbf{B}_{0}$ and the possible parallel wavelength $\lambda_{z0}$ are
restricted by the loop length $L_z$, $\lambda_{z0}=2\pi /k_{z0}\leq L_z$
(wavenumber $k_{z0}\gtrsim 2\pi /L$).

For the wavelengths within the mentioned above limits, a large spectral
flux, and hence a strong plasma heating, can be established by the strong
turbulence driven at the critically-balanced anisotropy $\alpha k_{\perp
0}/k_{z0}=$ $\alpha \lambda_{z0}/\lambda_{\perp 0}=$ $v_{A}/v_{k0}$. The
corresponding energy flux injected in the unit volume is $\epsilon
_{cor}\approx \alpha_{0}\rho v_{k0}^{3}/\lambda_{\perp 0}$. Assuming that
the turbulent velocity at injection $v_{k0}$ is observed as the nonthermal
velocity, $v_{k0}\approx v_{\mathrm{nt}}$, and taking from \citet{Xie2017}
$v_{\mathrm{nt}}=30$ km~s$^{-1}$, magnetic field $B=30$~G, and density
$n_{e}=2\times 10^{9}$ cm$^{-3}$, we obtain the energy flux $\epsilon
_{cor}\sim \alpha_{0}\rho v_{\mathrm{nt}}^{3}/l_{\perp}\sim $ $3\times
10^{-4}\left( l_{\perp}/L_{\perp}\right) ^{-1}$ erg cm$^{-3}$ s$^{-1}$.
For sufficiently small $l_{\perp}\lesssim 0.1L_{\perp}$, the energy flux
$\epsilon_{cor}\gtrsim $ $3\times 10^{-3}$ erg cm$^{-3}$ s$^{-1}$ is enough
to heat typical coronal loops. The corresponding parallel wavelengths at
injection are $\lambda_{z0}\sim \left( \alpha v_{\mathrm{nt}}/v_{A}\right)
^{-1}\lambda_{\perp 0}\lesssim 0.5L_z$. Therefore, the turbulent cascade
and related plasma heating can be effective if the perpendicular length
scales of the loop substructures are about 10 times smaller than the loop
width, which implies that the loops should be structured more than was
required by previous turbulent heating models.

\section{Conclusions}

In the framework of compressional MHD, we studied numerically the spectral
transport produced by the collisions between counter-propagating Alfv\'{e}n
waves. The initial two waves are linearly polarized in two orthogonal planes
and their cross-field profiles vary normally to their polarization planes.
Polarization and spectral characteristics of the perturbations generated
after single and multiple collisions between such AWs are analyzed in
detail. The main properties of the resulting spectral transfer are as
follows:

\begin{itemize}
\item the perturbations generated by AW collisions have smaller scales than
the original waves, which supports turbulence scenarios based on the direct
turbulent cascade generated by AW collisions;

\item we observed two regimes of the AW interaction: the first one is
typical for the case of strong turbulence, and the second one is governed by
a different mechanism;

\item the spectral transfer generated by the AW collisions is strongly
suppressed at the scales satisfying the classic critical balance condition (\ref{old})
of incompressional MHD, which makes the turbulence weak at these
scales;

\item the strong turbulence is re-established at significantly smaller
perpendicular scales satisfying the modified critical balance condition (\ref{new});
\end{itemize}

We used these properties to re-evaluate the turbulent heating of the solar
coronal loops. The main conclusion is that the turbulent cascade can heat
the loop plasma provided the loop is structured and the characteristic
widths of the loop sub-structures are more than 10 times smaller than the
loop width.

\bibliographystyle{aa}
\bibliography{shestov_voitenko}

\end{document}